# DJI drone IDs are not encrypted

Conner Bender
The University of Tulsa
`conner@utulsa.edu`

*Abstract*—Drones are widely used in the energy, construction, agriculture, transportation, warehousing, real estate and movie industries. Key applications include surveys, inspections, deliveries and cinematography. With approximately 70-80% of the global market share of commercial off-the-shelf drones, Da-Jiang Innovations (DJI), headquartered in Shenzhen, China, essentially monopolizes the drone market. As commercial-off-the-shelf drone sales steadily rise, the Federal Aviation Administration has instituted regulations to protect the federal airspace. DJI has become a pioneer in developing remote identification technology in the form of drone ID (also known as AeroScope signals). Despite claims from the company touting its implementation of drone ID technology as "encrypted" yet later being proved incorrect for the claim, it remains a mystery on how one can grab and decode drone IDs over the air with low-cost radio frequency hardware in real-time. This research paper discusses a methodology using radio software and hardware to detect both Enhanced Wi-Fi and OcuSync drone IDs, the three types of drone ID packet structures and a functioning prototype of a DJI OcuSync detection system equipped with two HackRF Ones.

## I. INTRODUCTION

DJI originally used incorrect verbiage when describing its remote identification technical specifications as "encrypted." In April 2022, DJI originally told The Verge that its drone IDs are encrypted. The company was later proved incorrect after security researcher Kevin Finisterre showed via Twitter that a Mavic Mini SE drone ID, which it broadcasted over Enhanced Wi-Fi, can be detected in cleartext [15].

This research paper builds upon the work discovered by Kevin Finisterre by further validating the thesis that DJI drone IDs are not encrypted in both Enhanced Wi-Fi and OcuSync protocols. While the means of OcuSync demodulation are kept secret outside company walls, the methods to crack it do not break any encryption techniques. True cryptographic encryption requires a key to decrypt information, and no keys are required for OcuSync demodulation. This research walks through the demodulation steps, which were provided by open-source source avenues.

In addition, this research paper discusses the three types of drone ID packet structures: license, flight information version 1 and flight information version 2. These packet types are recognized by a working DJI OcuSync detection demonstrated in this research. The detection system is equipped with two HackRF Ones and runs a web application responsible for displaying real-time detection data. Lastly, this research shows how to achieve DJI Enhanced Wi-Fi detection and extending such detection capabilities to non-DJI Wi-Fi drones such as Parrot.

## II. BACKGROUND

Civilian unmanned aerial vehicles, also known as drones, are commercial-off-the-shelf products that have gained popularity over the years. The Federal Aviation Administration estimates that commercial-off-the-shelf drone sales rose from 2.5 million in 2016 to 7 million in 2020 [12].

The Federal Aviation Administration is charged by the Code of Federal Regulations to be the ultimate arbiter in civil aviation regulation. Specifically, the Federal Aviation Administration oversees Title 14, Chapter 1 in the Code of Federal Regulations, which has two parts that cover drones. Part 107 contains regulations that govern small drones [11]. Part 89 contains remote identification requirements for small drones [10].

Due to the increased operation of drones, the Federal Aviation Administration has instituted regulations to protect the federal airspace. The regulations cover documented authorization, flight operating limitations and remote identification requirements. A key regulation is that, after September 16, 2023, drones that do not comply with the remote identification reporting requirements may not be operated in federal airspace [8].

Remote identification, which is implemented by drones broadcasting vehicle and flight data once every second, enables the Federal Aviation Administration, law enforcement and other federal agencies to properly identify vehicles and their ground stations. Remote identification of a drone is implemented using messages containing specific data elements that are emitted by the vehicle when it is turned on. The messages must be broadcasted once per second [10].

Table 1 shows the remote identification elements and performance requirements from Section 305 and Section 310 in Title 14, Chapter 1, Part 89 in the Code of Federal Regulations. The remote identification elements are prioritized so that aviation and law enforcement authorities can quickly and accurately identify persons who misuse drones.

The drone serial number and session identifier are unique tags set by the manufacturer. Manufacturers may incorporate the serial number or session identifier, or both, in broadcast messages. The longitude and latitude data points of the controller must be encoded cryptographically before broadcasting. The controller altitude must be the most precise measurement reported; the measurement must have a margin of error no more than 15 feet. The timestamp in the message must be synchronized within one second with all other data elements from the time of broadcast. The broadcast message must also have some form of error correction embedded.

The elements are emitted by an internal broadcasting module in the drone. However, if a manufacturer decides to create its own external broadcasting module, the module does not have to report controller location and altitude information. The Federal Aviation Administration final authority states that a drone may not take off if the elements in Table 1 are not included in its broadcast messages.

TABLE I. REMOTE IDENTIFICATION REQUIREMENTS OF DRONES.

| Elements | Performance |
|---|---|
| Drone serial number and/or session identifier | Rate of one message per second |
| Controller latitude and longitude | ±100 feet with 95% probability |
| Controller geometric altitude | ±15 feet with 95% probability |
| Drone latitude and longitude | ±100 feet with 95% probability |
| Drone geometric altitude | ±150 feet with 95% probability |
| Drone velocity | Rate of one message per second |
| Timestamp | Synchronized with all other elements |
| Drone emergency status | On/Off |

### A. Da-Jiang Innovations

With approximately 70-80% of the global commercial off-the-shelf drone business, Da-Jiang Innovations (DJI), headquartered in Shenzhen, China, essentially monopolizes the drone market [21]. DJI releases several new drone models every year. The models are typically used in three drone applications: video production, enterprise activities and agricultural solutions.

A DJI drone currently employs one of two proprietary communications protocols, Enhanced Wi-Fi and OcuSync [10]:

1. *Enhanced Wi-Fi Protocol:* The Enhanced Wi-Fi protocol is used by older DJI Spark and Mavic Air models. The protocol transmission range is limited to visual line of sight.

2. *OcuSync Protocol:* The OcuSync protocol is used by the DJI Mavic series, Air series and Mini series of drones. This new DJI protocol, which leverages software-defined radio technology, has a protocol transmission range of approximately 2.5 miles.

The Lightbridge protocol was the first protocol DJI implemented and was used in older DJI Phantom series, Inspire series and Matrice series of drones. Newer DJI drones do not implement Lightbridge, which is why it is not discussed in this research.

DJI also markets drone monitoring systems that leverage the remote identification functionality mandated by the Federal Aviation Administration to ensure that drones are operated safely. In 2017, DJI released AeroScope monitoring systems that can identify most DJI drones (Lightbridge, Enhanced Wi-Fi and OcuSync drones). Fig. 1 shows a stationary unit and a mobile unit. The latest DJI AeroScope drone monitoring systems can identify up to 50 DJI drone models [4]. DJI markets these systems as ideal for placement at or near airports, prisons, nuclear power plants, government facilities and critical infrastructure assets [5].

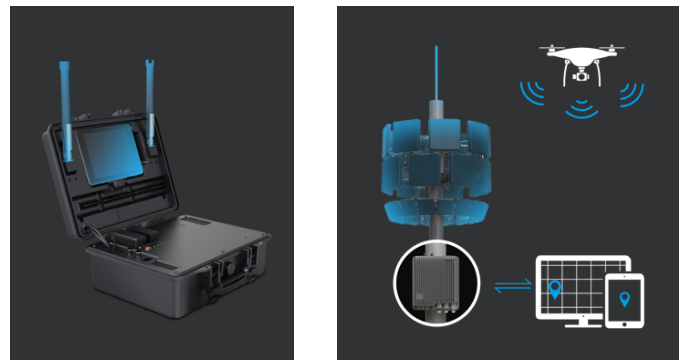

Fig. 1. DJI AeroScope stationary unit (left) and mobile unit (right).

A DJI AeroScope monitoring system captures and stores telemetry data of all the drones it monitors as well as their serial numbers and GPS data, return-to-home GPS data and controller (pilot) GPS data. The unique user identification code (UUID) associated with the purchaser of a drone is also recorded. DJI has confirmed that the UUID ties a drone pilot to a registered DJI user account [5]. Thus, DJI can query UUIDs to obtain pilot names, email addresses, even credit card numbers and related data. Law enforcement agencies requesting UUID information may be granted personal identifiable information regarding the drone owner.

*B. DJI Remote Identification for Enhanced Wi-Fi*

Department 13, a company specializing drone countermeasures, discovered operational details about drone remote identification packets by examining an accidental release of DJI's Mavic Pro drone firmware. The discovery is documented in a white paper that provides the only publicly available information about DJI remote identification frames sent with Enhanced Wi-Fi drones [8].

DJI drone remote identification packets are incorporated in IEEE 802.11 Wi-Fi beacon management frames. The 802.11 Wi-Fi beacon management frames are designed to announce the existence of vehicles [13]. Fig. 2 shows the 802.11 beacon management frame structure Enhanced Wi-Fi drones emit. The frame control field specifies the frame type and subtype; setting the type to 0 and subtype to 8 informs the network that the frame is a beacon. The device address, sender address and basic service set identifier (BSS ID) fields are arbitrary. The frame body contains various information elements. The information elements provide the network with detailed information about access points. The service set identifier (SSID) field, supported rates field and current channel field are included in the frame.

```
v16 = flight_info_len;
if (flight_info_len)
{
    bcnbuf[0] = 0xDDu;
    *(_WORD *)&bcnbuf[2] = 0x3726;
    *(_WORD *)&bcnbuf[4] = 0x5812;
    *(_WORD *)&bcnbuf[6] = 0x1362;
    memcpy(&bcnbuf[8], flight_info, flight_info_len);
    bcnbuf[1] = v16 + 6;
}
ath6kl_wmi_set_appie_cmd(v4[17], v12->fw_vif_idx, ...
0, bcnbuf, bcnbuf[1] + 2);
dji_ie_on = 1;
return v5;
```

Fig. 3. Generation of DJI drone ID Enhanced Wi-Fi beacon frames.

### III. OCUSYNC DRONE IDENTIFICATION

In 2022, an open-source effort was initiated to demodulate DJI OcuSync drone ID signals [18]. The repository revealed that OcuSync drone IDs are loosely based on some Long Term Evolution (LTE) cellular standards. This means any software defined radio (SDR) capable of sampling up to 15,360,000 samples per second (the LTE sample rate) up to a bandwidth of 10 MHz can capture OcuSync drone IDs.

Fig. 4 shows a GNU Radio flow graph leveraging a LimeSDR to see DJI drone ID signals in real-time (via a fosophor sink). Any SDR capable of sampling 32-bit floating point IQ data (.fc32 file) up to 15.35 MSPS can use this GNU Radio flow graph and capturing drone ID signals. Fig. 5 shows the waterfall output from the GNU radio flow graph with a drone ID highlighted.

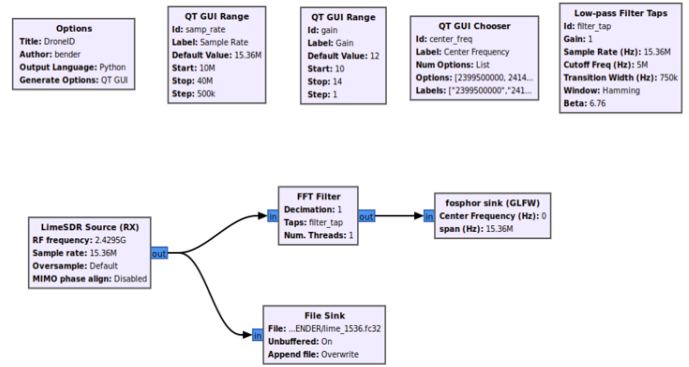

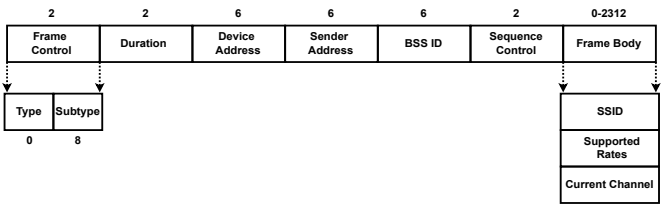

Fig. 2. 802.11 Wi-Fi beacon management frame structure.

Appended to an 802.11 Wi-Fi beacon management frame is a DJI drone payload. Fig. 3 discovered in the decompiled code of the `ath6kl_usb.ko` firmware kernel module shows the structure of the payload [8]. The kernel module directly controls the Atheros wireless network adapter in the drone. In the code snippet, the payload preamble is a unique series of eight bytes concatenated (via `memcpy`) to the `flight_info` packet. After the `dji_ie_on` parameter is set, the drone begins to emit a 76-byte remote identification packet via the Atheros wireless network adapter.

Fig. 4. GNURadio flow graph for LimeSDR to capture drone IDs.

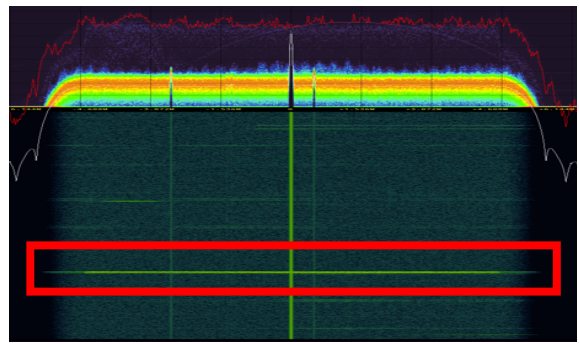

Fig. 5. Waterflow showing capture of DJI drone ID (highligthed).

With a capable SDR, a successful capture must be tuned to center frequencies broadcasted by drone ID signals. Table 2 shows center frequencies in the 2.4 GHz and 5.8 GHz frequency bands that drone IDs are found on. This data was collected by leaving several DJI Mavic models with the power on. The 2.4 GHz frequency band ranges from 2399.5 MHz to 2474.5 MHz, and the 5.8 GHz frequency band ranges from 5741.5 MHz to 5831.5 MHz. It was also discovered that even if a user forces the DJI OcuSync communications downlink to be 2.4 GHz or 5 GHz with the DJI GO smartphone application, DJI drone ID signals do not adhere and continue to broadcast out-of-band from the communications link.

TABLE II.  2.4 GHz CENTER FREQUENCIES AND 5.8 GHz CENTER FREQUENCIES FOR DJI DRONE IDs.

| 2.4 GHz Frequency Band | 5.8 GHz Frequency Band |
|---|---|
| 2399.5 MHz | 5741.5 MHz |
| 2414.5 MHz | 5756.5 MHz |
| 2429.5 MHz | 5771.5 MHz |
| 2444.5 MHz | 5786.5 MHz |
| 2459.5 MHz | 5801.5 MHz |
| 2459.5 MHz | 5816.5 MHz |
|  | 5831.5 MHz |

A signal broadcasts on a frequency until 12-20 drone IDs emit before hopping. Fig. 6 shows a sample DJI Mavic Pro drone ID signal. A drone ID signal is approximately 600 milliseconds in duration.

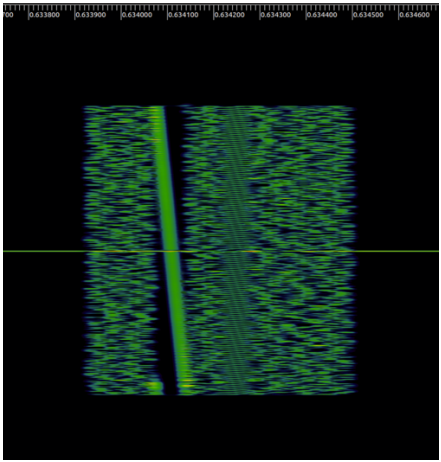

Fig. 6. A sample DJI Mavic Pro drone ID signal captured on 2429.5 MHz.

A drone ID signal contains nine OFDM (Orthogonal Frequency-Division Multiplexing) symbols. Sometimes, drone ID signals contain eight OFDM symbols. However, demodulation only needs eight OFDM symbols as the first symbol in a nine OFDM system can be discarded.

The first and last OFDM symbols are sized 80 and the middle symbols are sized 72 (e.g., ([80, 72, 72, 72, 72, 72, 72, 72, 80]). Once captured, DJI drone ID signals can be demodulated using digital signal processing. The demodulation steps are provided in the dji_droneid GitHub repository and are the following [18]:

1. Identifying the start of a drone ID
2. Creating a low-pass bandwidth filter
3. Applying a coarse frequency offset correction
4. OFDM symbol extraction (sans cyclic prefixes)
5. Measuring channel impulse rate
6. Quantizing Quadrature Phase Shift Key (QPSK) into bits
7. Descrambling bits
8. Turbo decoder and rate matcher
9. Deframing bytes

The GitHub repo is written in MATLAB/Octave code and used a Ettus B205-mini SDR ($1345) for research. In this research paper, the demodulation steps were converted into Python. The proceeding algorithms in this paper outline each step in Pythonic pseudocode.

### A. Start of Drone ID

Alg. 1 generates Zadoff-Chu (ZC) sequences with a root index and sequence length [1]. In a drone ID, there are two OFDM symbols with ZC sequences, symbol 4 and 6, in it. The rootIndex for OFDM symbol 4 is 600 and the rootIndex for OFDM symbol 6 is 147. The seqLen is 601 because the formula only works for an odd number of samples. The middle sample (300) is removed after computation. The sequence is then applied to the data carriers (in buffer). The buffer is shifted to have the zero value placed in the center and an inverse Fourier transformation occurs. The result of the algorithm yields a 600-long ZC sequence shifted with a root index (zadoffChuSeq).

---

**Algorithm 1:** Generation of Zadoff-Chu sequence.

**Input:** rootIndex: Root index of Zadoff-Chu symbol
**Input:** seqLen: Length of Zadoff-Chu sequence
**Output:** zadoffChuSeq: Zadoff-Chu sequence
dataCarrierIndices ← $[i = 212 \ldots 813, i \neq 512]$
n ← $[i = 0 \ldots 600]$
seq ← $e^{-1j \times \pi \times \text{rootIndex} \times n \times (n + 1 + 2. \times 0)/\text{seqLen}}$
seq ← delete(seq, 300)
buffer ← zeros(1024)
buffer [dataCarrierIndices] ← seq
zadoffChuSeq ← invFFT(shiftToCenter(buffer))

Alg. 2 performs a normalized cross correlation that finds the ZC sequence in OFDM symbol 4 (zc4) among 32-bit floating point IQ data (iqData). This is accomplished with the NumPy correlate function. The next step is selecting the largest peak found in crossCorrelation (minimum and maximum parameters vary based on signal strength) using the findPeaks function. After the greatest peak is identified, the start of the drone ID burst (startBurst) can be found by backtracking four OFDM symbols in length. A clean drone ID burst is trimmed from startBurst up until burstDuration.

**Algorithm 2:** Normalized cross correlation.
  **Input:** iqData: 32-bit floating point IQ data
  **Output:** burst: Drone ID burst
  zc4 ← zadoffChuSeq(600, 601)
  crossCorrelation ← | correlate(iqData, zc4) $|^2$
  peak ← findPeaks(crossCorrelation, height=(1e5,1e6))
  fftSize ← 1024
  zcOffset ← 80 + (72 × 3) + (fftSize × 3)  # burst start
  startBurst ← peak – zcOffset
  burstDuration ← (80 × 2) + (72 × 7) + (fftSize × 9)
  burst ← burst[startBurst:startBurst+burstDuration]

### B. Low-pass Bandwidth Filter

Alg. 3 performs a low-pass bandwidth filter on a drone ID burst. The filter function is from the signal package. The filter window firWin is designed to fit the drone ID burst bandwidth (bw) and sample rate (sampRate) at a length of n, which is 51. The filterTaps object is a low-pass bandwidth filter that is applied to the drone ID burst, resulting in filteredBurst.

**Algorithm 3:** Applying a low-pass band filter.
  **Input:** burst: Drone ID burst
  **Output:** filteredBurst: Filtered drone ID burst
  n ← 51
  bw ← 10e6
  sampRate ← 15.36e6
  filterTaps ← firWin(n, bw/sampRate)
  filteredBurst ← digitalFilter(filterTaps, burst, axis=0)

Fig. 7 shows the graphical result of the low-pass bandwidth filter. The graph plotted the magnitude (squared) of the drone ID in a log scale.

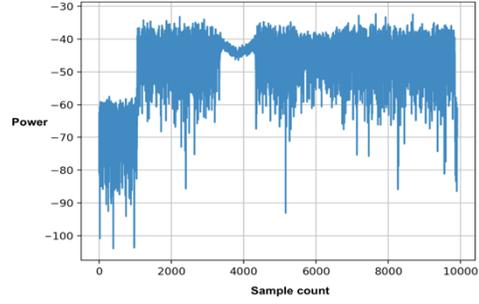

Fig. 7. DJI drone ID burst with low-pass bandwidth filter.

### C. Coarse Frequency Offset Correction

Alg. 4 shows the coarse frequency offset calculation. The coarse frequency offset is caused by a frequency offset from the SDR, which is minor in comparison to an actual frequency offset. The coarse frequency offset is calculated by inspecting the second OFDM symbol—specifically, the symbol cyclic prefixes. The second OFDM symbol begins with a cyclic prefix and ends with an inverted cyclic prefix. The cp variable denotes the first cyclic prefix in the second OFDM symbol and the copy variable denotes the second cyclic prefix. In NumPy, a dot product operation is obtained by conjugating cp before multiplying it with copy. The result is the sum of all the elements along axis 0. The offsetRadians is generated on the complex plane in radians, which is inversely applied back to the burst to perform a coarse frequency offset correction in the drone ID burst.

**Algorithm 4:** Correction of the coarse frequency offset.
  **Input:** burst: Filtered drone ID burst
  **Output:** newBurst: Frequency offset corrected burst
  fftSize ← 1024
  cp ← burst [1104:1176] # $1^{st}$ cyclic prefix
  copy ← burst[2128:2200]  # $2^{nd}$ cyclic prefix
  offsetRadians ← angle(sum($\bar{}$start × end, axis=0)) / fftSize
  newBurst ← burst × $e^{1j \times \text{-offsetRadians} \times [1\ldots\text{length(burst)}+1]}$

### D. OFDM Symbol Extraction

Alg. 5 shows the time and frequency domains generated from the drone ID burst. The process for extracting OFDM symbols in the drone ID burst involves stripping cyclic prefixes and then converting remnants into the time and frequency domains. There are nine iterations that start at the end of each cyclic prefix in an OFDM symbol. Before the algorithm beings, the burst is already in the form of a time domain, so each symbol is stored in the corresponding timeDomain array row. The timeDomain is then converted to the frequency domain (freqDomain) by computing a Fourier transformation, followed by shifting the zero-frequency component to the center. The final array is stored in the corresponding freqDomain array row.

**Algorithm 5:** Generation of time and frequency domains.
   **Input:** burst: Drone ID burst
   **Output:** timeDomain: Time domain of symbols
   **Output:** freqDomain: Frequency domain of symbols
   prefixes ← [72, 80, 80, 80, 80, 80, 80, 80, 72]
   fftSize ← 1024
   freqDomain ← zeros(length(prefixes), fftSize)
   timeDomain ← zeros(length(prefixes), fftSize)
   offset ← 0
   **for** $i$ in range(length(prefixes)):
      offset ← prefixes[$i$] + offset
      timeDomain[$i$,:] ← burst[offset:offset+fftSize]
      freqDomain[$i$,:] ← fftShift(fft(timeDomain[$i$,:]))
      offset ← fftSize + offset
   **end**

### E. Channel Impulse Response

Alg. 6 shows the channel impulse response, which is measured by calculating the average walking phase offset. The channel impulse response represents the distortion of a signal [1]. Alg. 1 generates both ZC sequences found in OFDM symbols 4 and 6 (zc4 and zc6). Each ZC sequence is converted to the frequency domain by a Fourier transformation. The Golden references for each ZC sequence are stored in channel1 and channel2 for OFDM symbols 4 and 6, respectively. The channel estimation (est) is derived from channel1. The elements that are not data carriers are discarded. The average phase offset (phaseOffset) for each symbol is calculated by computing the angle of channel1 and channel2 and summing all the elements together. That sum is then divided by the number of data carriers (600). The final phaseOffset value is the average of both channel1 and channel2.

**Algorithm 6:** Measurement of channel impulse response.
   **Output:** phaseOffset: Average walking phase offset
   **Output:** est: Channel estimation
   dataCarrierIndices ← [$i$ = 212…813, $i \neq 512$]
   zc4 ← fftShift(fft(zadoffChuSeq(600, 601)))
   zc6 ← fftShift(fft(zadoffChuSeq(147, 601)))
   channel1 ← $zc4$/freqDomain[3,:]  # OFDM symbol 4
   channel2 ← $zc6$/freqDomain[5,:]  # OFDM symbol 6
   channel1 ← channel1[dataCarrierIndices]
   channel2 ← channel2[dataCarrierIndices]
   est ← channel1
   channel1Phase ← sum(angle(channel1), axis=0)/600
   channel2Phase ← sum(angle(channel2), axis=0)/600
   phaseOffset ← (channel1Phase - channel2Phase)/2

### F. Quantize QPSK into Bits

Alg. 7 shows the process of demodulating QPSK into constellation mappings (bits). The algorithm equalizes the frequency domain to only include data carriers, adjusting the sample to the previously calculated phaseOffset. The absolute phase offset is calculated from multiplying phaseOffset by the distance each OFDM symbol is from the symbol that was used for equalization. Because the phase offset was calculated between both ZC sequences (which are in OFDM symbols 4 and 6), the phaseOffset is directly applied to the OFDM symbol 5. The algorithm then loops through dataCarriers and converts the complex samples (that represent QPSK constellation points) into bits.

**Algorithm 7:** Quantization of QPSK to bits.
   **Input:** est: Channel estimation
   **Input:** phaseOffset: Average walking phase offset
   **Output:** demodBits: Quantized bits from drone ID
   carrierIndices ← [$i$ = 212…813, $i \neq 512$]
   demodBits ← zeros(9, 1200)
   **for** $a$ in length(bits):
      dataCarriers ← freqDomain[$a$, carrierIndices] × est
      dataCarriers ← $e^{1j \times \text{-phaseOffset} \times (a-4)}$
      offset ← 0
      quantizedBits ← zeros(1200)
      **for** $b$ in length(dataCarriers):
         sample ← dataCarriers[$b$]
         **if** *real(sample) > 0* and *imag(sample) > 0* **then**
            bits ← [0,0]
         **end**
         **elif** *real(sample) > 0* and *imag(sample) < 0* **then**
            bits ← [0,1]
         **end**
         **elif** *real(sample) < 0* and *imag(sample) > 0* **then**
            bits ← [1,0]
         **end**
         **elif** *real(sample) < 0* and *imag(sample) < 0* **then**
            bits ← [1,1]
         **end**
         **else**
            bits ← [0,0]
         **end**
         quantizedBits[offset:offset+2] ← bits
         offset ← offset+2
      **end**
      demodBits[$a$,] ← quantizedBits
   **end**

Table 3 shows the QPSK constellation mapping. The algorithm saves each quantized data carrier into demodBits.

TABLE III.    QPSK MODULATION MAPPING.

| b(*i*), b(*i*+1) | I | Q |
|---|---|---|
| 00 | $1/\sqrt{2}$ | $1/\sqrt{2}$ |
| 01 | $1/\sqrt{2}$ | $-1/\sqrt{2}$ |
| 10 | $-1/\sqrt{2}$ | $1/\sqrt{2}$ |
| 11 | $-1/\sqrt{2}$ | $-1/\sqrt{2}$ |

### G. Descramble Bits

Alg. 8 descrambles demodulated drone ID bits into complex values. The two initial values for descrambling are hardcoded polynomial values: lsfrX1 (whose value is outlined in 3GPP 36.211 7.2) and lsfrX2 (`0x12345678` for drone IDs) [1]. The variable $n_c$ is defined in the LTE standards as 1600. The first loop generates the m-sequence for lsfrX1. The second loop generates the m-sequence for lsfrX2. The third loop generates the resulting Gold sequence (goldSeq). The final operation is performs a bitwise XOR $\oplus$ operation between the OFDM symbols (2, 3, 5, 7, 8 and 9) and the goldSeq.

---

**Algorithm 8:** Descrambling bits into complex values.

**Input:** demodBits: Quantized bits from drone ID burst
**Output:** descBits: Complex values from descrambling

$$\text{lfsrX1} \leftarrow \begin{bmatrix} 1 & 0 & 0 & 0 & 0 & 0 & 0 & 0 \\ 0 & 0 & 0 & 0 & 0 & 0 & 0 & 0 \\ 0 & 0 & 0 & 0 & 0 & 0 & 0 & 0 \\ 0 & 0 & 0 & 0 & 0 & 0 & 0 & 0 \end{bmatrix}$$

$$\text{lfsrX2} \leftarrow \begin{bmatrix} 0 & 0 & 0 & 1 & 1 & 1 & 1 & 0 \\ 0 & 1 & 1 & 0 & 1 & 0 & 1 & 0 \\ 0 & 0 & 1 & 0 & 1 & 1 & 0 & 0 \\ 0 & 1 & 0 & 0 & 1 & 0 & 0 & 0 \end{bmatrix}$$

finalSeqLen ← 7200
$n_c$ ← 1600
x1 ← zeros($n_c$ + finalSeqLen + 31)
x2 ← zeros($n_c$ + finalSeqLen + 31)
goldSeq ← zeros(finalSeqLen, type='int8')
x1[0:31] ← lfsrX1
x2[0:31] ← lfsrX2
**for** *i* in length(finalSeqLen + $n_c$)
    x1[*i*+31] ← (x1[*i*+3] + x1[*i*]) % 2
**end**
**for** *i* in length(finalSeqLen + $n_c$)
    x2[*i*+31] ← (x2[*i*+3] + x2[*i*+2] + x2[*i*+*i*] + x2[*i*]) % 2
**end**
**for** *i* in length(finalSeqLen)
    goldSeq[*i*] ← (x1[*i*+$n_c$] + x2[*i*+$n_c$]) % 2
**end**
demodBits ← demodBits[[1, 2, 4, 6, 7, 8],:]
descBits ← demodBits $\oplus$ goldSeq

---

### H. Turbo Decoder and Rate Matcher

The turbo decoder is implemented using the `turbofec` library [19]. The 7200 descrambled bits from Alg. 8 are passed into the turbo code remover C++ program, which was provided by the `dji_droneid` GitHub repository [18]. The program sets up the necessary structures and buffers to interface with `turbofec` for turbo decoding and rate matching logic. The program outputs decoded data when the CRC-24 check returns `0x00`. Else, it outputs the calculated CRC-24 error. Poor SDR recordings, interference and frequency offsets often attribute to failed decoding calculations.

### I. Deframe Bytes

There are three types of drone ID packets this research has identified: license, flight information version 1 and flight information version 2.

Fig. 8 shows the structure of a license plate packet. License packets have a packet type of `0x11`. After the packet type comes a serial number of the detected drone. Following the drone serial number are the custom license and flight plan values provided by the DJI GO smartphone application user.

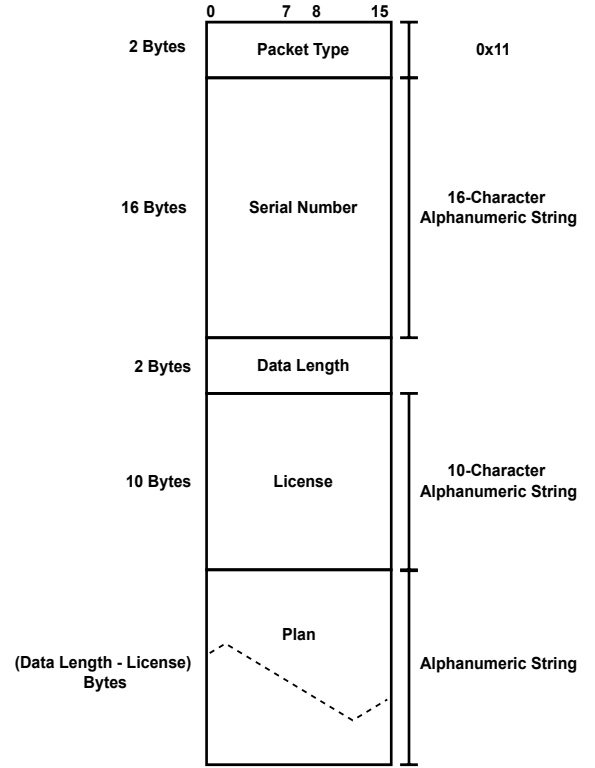

Fig. 8. Drone ID license packet.

Fig. 9 shows the structure of a version 1 flight information packet. Version 1 packets have a packet type of `0x1001`. The state information varies depending if the drone is operating properly (e.g., motors on, in air, home point is set, etc.) [8]. After the state information comes a serial number of the detected drone. The drone GPS coordinates, altitude, height, x speed, y speed, z speed, pitch angle, roll angle, yaw angle of the drone and return-to-home GPS coordinates are arranged sequentially. Each GPS coordinate (longitude and latitude) is packed into two bytes using the following computation:

$$\frac{GPS\ Coordinate}{180} \times \pi \times 10^7$$

Next in the packet are the model field that specifies the product type of the drone and the UUID, an 18-character string identifier that ties the unmanned aerial vehicle to a DJI user account.

Fig.10 shows the structure of a version 2 flight information packet. Version 2 packets have a packet type of 0x1002. The rest of the packet follows the same arrangement as a version 1 packet except there are no pitch angles or roll angles—only yaw angles. After the yaw angle comes the pilot GPS clock, which measures the number of milliseconds since epoch (January 1st, 1970), to assess phone GPS accuracy. The pilot GPS coordinates, return-to-home GPS coordinates, model, UUID length and UUID are arranged sequentially.

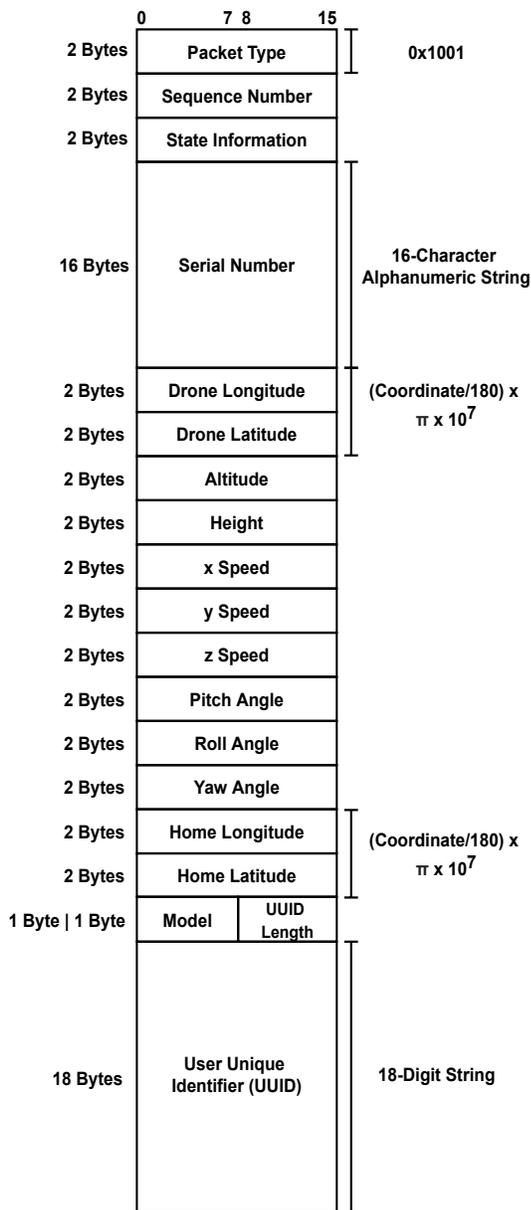

Fig. 9. Drone ID flight information packet (version 1).

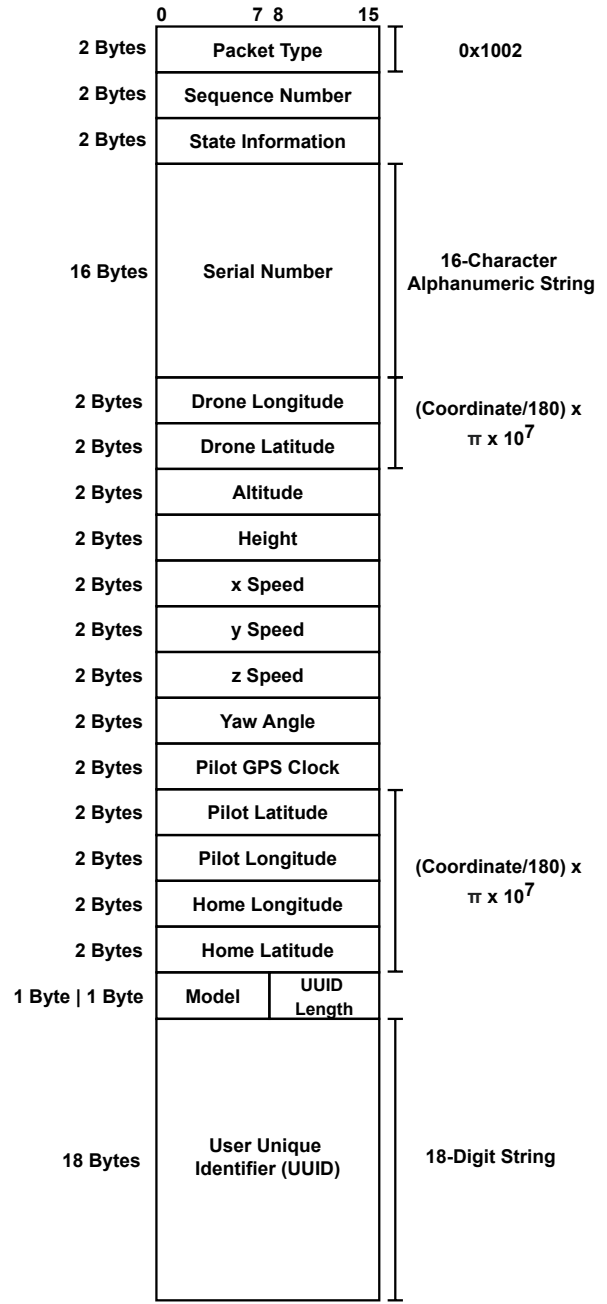

Fig. 10. Drone ID flight information packet (version 2).

Alg. 9 shows the conversion between aerodynamic angles to quantity values. Pitch, roll and yaw angles can be converted to quantities using a set of conditionals. If the angle is 0, the quantity yields 0. If the angle is less than 0, adding 180 to the angle to yields the quantity. If the angle is greater than 0 but less than 180, modular dividing the angle by 180 yields the quantity. Else, adding 180 to the angle to yields the quantity.

---
**Algorithm 9:** Convert aerodynamic angles to quantities.
    **Input:** angle: Pitch, roll or yaw angle
    **Output:** quantity: Pitch, roll or yaw
    **if** angle is 0 **then**
        quantity ← 0
    **end**
    **elif** angle < 0 **then**
        quantity ← angle + 180
    **end**
    **elif** angle > 0 and angle < 180 **then**
        quantity ← angle % 180
    **end**
    **else**
        quantity ← angle + 180
    **end**

---

In the event of only detecting a license packet, which contain a serial number and no model information, there is a way to predict the model of the drone by only looking at serial number prefixes (universal three-string constants). Table 4, data derived from the Federal Aviation Administration Aircraft Inquiry Database and DJI Service Request and Inquiry website, shows DJI models corresponding to serial number prefixes [7, 9]. The table reveals all the drones equipped with remote identification capabilities. Also, the table shows all possible AeroScope IDs (value in the model field) for both versions of flight information packets. The AeroScope IDs were found on a DJI storage domain [4]. All the AeroScope IDs are associated with DJI models, except for AeroScope ID 240 which is associated with a non-DJI Yuneec H480 drone. Yuneec is a DJI competitor, so the AeroScope ID associated with the Yuneec H480 model may be present because DJI was attempting to detect Yuneec drones and/or DJI has plans to acquire Yuneec.

TABLE IV. DJI MODELS, SERIAL NUMBER PREFIXES AND KEYS.

| Model | Prefix | AeroScope ID |
|---|---|---|
| Inspire 1 | 041, W21 | 1 |
| Phantom 3 Series | 0JX | 2 |
| Phantom 3 Series Pro | P76 | 3 |
| Phantom 3 Std | 03Z, P5A | 4 |
| M100 | M02 | 5 |
| ACEONE | – | 6 |
| WKM | – | 7 |
| NAZA | 061 | 8 |
| A2 | 061 | 9 |
| A3 | 067 | 10 |
| Phantom 4 | 07D, 07J, 0AX, 0HA, 189 | 11 |
| MG1 | 05Y | 12 |
| M600 | M64 | 14 |
| Phantom 3 4k | P7A | 15 |
| Mavic Pro | 08Q, 08R | 16 |
| Inspire 2 | 095, 09Y, 0A0 | 17 |
| Phantom 4 Pro | 0AX | 18 |
| N2 | – | 20 |
| Spark | 0AS, 0BM | 21 |
| M600 Pro | M80 | 23 |
| Mavic Air | 0K1, 0K4 | 24 |
| M200 | 0FZ | 25 |
| Phantom 4 Series | CE1 | 26 |
| Phantom 4 Adv | 0HA | 27 |
| M210 | 0N4 | 28 |
| M210RTK | 17U, 1DA | 30 |
| A3_AG | – | 31 |
| MG2 | – | 32 |
| MG1A | – | 34 |
| Phantom 4 RTK | 0UY, 0V2 | 35 |
| Phantom 4 Pro V2.0 | 11U, 11V | 36 |
| MG1P | 0YS | 38 |
| MG1P-RTK | 0YL | 40 |
| Mavic 2 | 0M6, 163 | 41 |
| M200 V2 Series | 17S | 44 |
| Mavic 2 Enterprise | 276, 29Z | 51 |
| Mavic Mini | 1SC, 1SD, 1SZ, 1WG | 53 |
| Mavic Air 2 | `1WN`, `3N3` | 58 |
| P4M | `1UD` | 59 |
| M300 RTK | `1ZN` | 60 |
| DJI FPV | `37Q` | 61 |
| Mini 2 | `3NZ`, `3Q4`, `5DX`, `5FS` | 63 |
| AGRAS T10 | `IEZ` | 64 |
| AGRAS T30 | `35P` | 65 |
| Air 2S | `3YT` | 66 |
| M30 | – | 67 |
| Mavic 3 | `F4Q`, `F45` | 68 |
| Mavic 2 Enterprise Adv | `298` | 69 |
| Mini SE | `4AE`, `4DT`, `4GM` | 70 |
| Mini 3 Pro | – | 73 |
| YUNEEC H480 | `YU1` | 240 |

## IV. DJI Drone Detection

There are ways to achieve real-time DJI drone detection without purchasing a DJI AeroScope. In fact, this is achieved using low-cost radio hardware.

### A. OcuSync Detection

A DJI OcuSync detection system runs effectively using "cheap" SDRs (under $500) such as HackRF Ones coupled with an Intel-based processor computer to process the Python script and turbo decoder program. The OcuSync drone ID demodulation steps and logic have already been discussed in this research paper but monitoring a DJI drone on the fly requires real-time detection capabilities with reliable radio frequency throughput. Rapid frequency hopping on known center frequencies (via Table 2) is also necessary to maximize success rates.

Alg. 2 can be outfitted to include a dynamic cross correlation that is constantly scanning for drone IDs. The `@njit` decorator in Numba, a high-performance Python compiler, can aid in faster calculations and yields better performance benchmarks when applied to the cross correlation function.

The tradeoff with low-cost SDRs (e.g., HackRF Ones) is unreliable crystal oscillators which occasionally invoke frequency offsets. These offsets can result in failed demodulations. However, there are ways to calculate a frequency offset at any juncture and apply a correction to a capture. The best solution found for a HackRF One is using the CellSearch program in the `LTE-Cell-Scanner` GitHub repository [20].

Alg. 10 leverages the CellSearch program to ping nearby LTE bands (stored in lteBand) from a range of frequencies: a frequency minimum (freqMin) to a frequency maximum (freqMax). The program can measure the crystal frequency error of the LTE frequency (crystalCorrection). The crystalCorrection is then transformed into a ppm number with a simple equation.

**Algorithm 10:** Correction of HackRF frequency offset.
  **Input:** freqBand: Nearby frequency band
  **Output:** ppm: HackRF crystal oscillator error
  lteBand ← {1: [2140, 2140.1]…103: [757.5, 757.6]}
  freqMin ← lteBand[freqBand][0]
  freqMax ← lteBand[freqBand][1]
  crystalCorrection ← CellSearch(freqMin, freqMax)
  ppm ← 1e6 × (1- crystalCorrection)

The ppm number can be inputted into the `-C` argument of the `hackrf_transfer` command (which interfaces with the HackRF to receive radio frequency data) to correct the frequency offset:

```
hackrf_transfer file.cs8 -f [CENTER_FREQUENCY]
-s 1536000000 -C [PPM]
```

The last step is converting `file.cs8`, which is a collection of complex 8-bit signed integer samples, into IQ data. The conversion can be performed with the following code snippet:

```
buffer = open(file.cs8, type=int8)
buffer = buffer [::2] + 1j × buffer[1::2]
```

Fig. 11 and 12 show the working prototype built for OcuSync drone ID detection. A refurbished Mac Mini, two HackRF Ones and two Altelix 2.4 GHz omni-directional antennas are used as the OcuSync drone ID detection system. The total cost is less than $1000, making it significantly cheaper than any detection system (including DJI AeroScope) on the market. One HackRF hops to a different center frequency on the 2.4 GHz frequency band and another HackRF hops to a different center frequency on the 5.8 GHz frequency band. Each capture consists of 1,500,000 complex 8-bit signed integers samples, which are enough samples to potentially contain a drone ID.

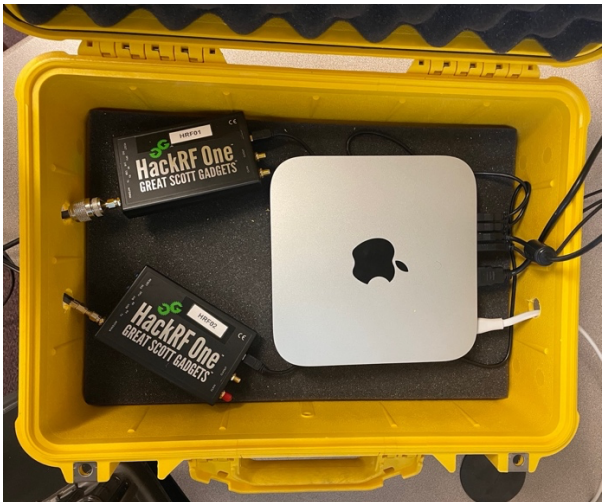

Fig. 11. Mac Mini and two HackRF Ones in a pelican case

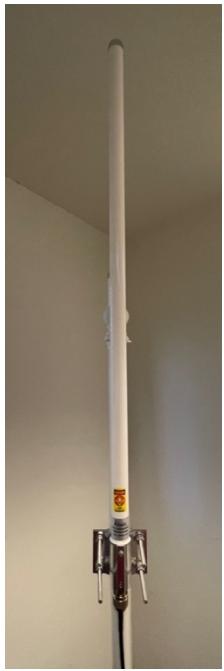

Fig. 12. Altelix Omni-directional 2.4 GHz antenna.

A Mavic Pro drone and a Mavic 2 Enterprise drone were used to test and verify the functionality of the OcuSync drone ID detection system. These drones were turned on and flying around within a few feet of the detection system.

Fig. 13 shows the script output of a detected Mavic 2 drone. In this instance, a drone ID version 2 packet is picked up and properly decoded. The script prints out the raw bytes of the deframed drone ID and JSON data of the packet data points. The script properly parses the serial number, GPS coordinates of the drone, altitude, height, speeds, yaw, GPS accuracy (pilot GPS clock), GPS coordinates of the pilot, GPS coordinates of the return-to-home location and UUID. The serial number and UUID have been concealed for the sake of privacy.

```
=================== Frame ===================
0000
0010
0020  62 02 02 00 03 00 F1 FF 36 1D 44 4E 5D 57 6E 01   b.......6.DN]Wn.
0030  00 00 DF 45 60 00 1F 73 00 FF 20 73 00 FF DF 45   ...E`..s.. s...E
0040
0050

{
  "model": "Mavic 2",
  "source_type": "OcuSync (SDR)",
  "packet_length": 94,
  "packet_type": "DroneID v2",
  "sequence_num": 119,
  "state_info": "0xf71f",
  "serial_num":             ,
  "drone_longitude": -95.94940313333159,
  "drone_latitude": 36.15195237683726,
  "altitude": 285,
  "height": 61.0,
  "x_speed": 0.02,
  "y_speed": 0.03,
  "z_speed": -0.15,
  "total_speed": 0.15427248620541512,
  "yaw": 254.78,
  "pilot_gps_clock": 1573423763.012,
  "pilot_longitude": -95.95751048613268,
  "pilot_latitude": 36.14987254004094,
  "home_longitude": -95.95750475655475,
  "home_latitude": 36.14987254004094,
  "uuid_len": 19,
  "uuid":          
}
```

Fig. 13. Script output of a detected Mavic 2 drone.

Fig. 14 shows the drone control panel web application built in Flask. The Flask application listens for JSON requests and displays nine detection attributes: model, serial number, UUID, altitude, height, yaw, drone location, pilot location and timestamp.

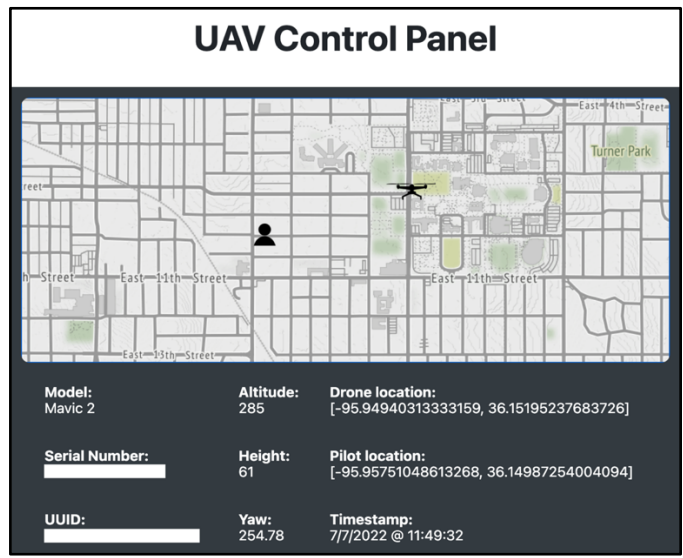

Fig. 14. Drone control panel showing detected Mavic 2 drone.

## B. Enhanced Wi-Fi Detection

A DJI Enhanced Wi-Fi detection system can be achieved by only using network adapters with monitor mode and custom clock rate capabilities.

The Department 13 white paper revealed that DJI drones use Atheros chips to broadcast beacon frames that contain drone IDs [8]. The beacon frames are emitted with a bandwidth of 5 MHz. This is typically not allowed for normal network interface cards; however, Atheros network interface cards are able to be clocked at half rate (10 MHz) or quarter rate (5 MHz) [2].

Fig. 15 shows a Qualcomm Atheros QCNFA435 M.2 WLAN/Bluetooth laptop Wi-Fi card that can detect Enhanced Wi-Fi drones. These cards be purchased from any online retailer for less than $30. An Atheros network interface card can also be used by Kismet, a wireless network device detector and sniffer that operates on Wi-Fi interfaces.

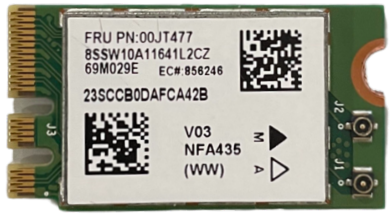

Fig. 15. Qualcomm Atheros Wi-Fi M.2 chip.

The following command allows Kismet to scan Wi-Fi channel 1 through 177 (consisting both 2.4 GHz and 5.8 GHz frequencies) at a 5 MHz bandwidth:

```
kismet -c adapter:channels = "1W5, 2W5,
3W5, 4W5, 5W5, 6W5, 7W5, 8W5, 9W5, 10W5,
11W5, 12W5, 13W5, 14W5, 140W5, 149W5,
153W5, 157W5, 161W5, 165W5, 169W5, 173W5,
177W5"
```

The Kismet control panel allows users to filter Wi-Fi devices based upon MAC address, advertised SSID or beacon vendor tag. The best way to detect DJI drones is by filtering individualized 802.11 vendor tags. IEEE keeps an open-source record of all registered company 802.11 vendor tags. DJI vendor tags are the following:

- 60-60-1F
- 34-D2-62
- 48-1C-B9

By extension, Kismet supports non-DJI drone detection such as Parrot. Parrot vendor tags are the following:

- 00-12-1C
- 90-03-B7
- A0-14-3D
- 00-26-7E

This Enhanced Wi-Fi detection proof-of-concept was not able to be tested against any DJI or Parrot drones.

## V. EXPERIMENTATION AND RESULTS

An experiment was conducted to assess the range capabilities the DJI OcuSync drone ID detection system has. The detection system was left running for two days and was able to detect three drones.

Table 5 shows the detection results from the experiment. The detection system was able to capture Mavic Pro flight information packet and a Mavic Mini license packet, both approximately 0.3 miles away. The furthest distance the detection system was able to find was a DJI Mavic 2, which was 0.75 miles away as shown in Fig. 13 and 14.

TABLE V. DETECTION RESULTS FROM EXPERIMENT.

| Model | Range | Packet Type |
| --- | --- | --- |
| Mavic Pro | ~0.30 miles | Flight information (v1) |
| Mavic Mini | ~0.30 miles | License |
| Mavic 2 | ~0.75 miles | Flight information (v2) |

## VI. CONCLUSIONS AND RECOMMENDATIONS

The proposed OcuSync drone ID detection system is superior in affordability, cost and flexibility in comparison to other solutions. It is the first demonstrated detection system that offers real-time detection of DJI OcuSync drone IDs using low-cost SDRs with robust packet dissection (license and flight information). The detection system is significantly cheaper than DJI AeroScope, which is marketed anywhere from $20,000 to $40,000, and can predict DJI drone models based on serial numbers. However, DJI AeroScope stationary units can detect drones up to 30 miles away whereas the detection system has only demonstrated a 0.75 mile detection range thus far.

Future work can be focused on increasing the range of the OcuSync drone ID detection system, conducting more experiments with the detection system and merging both Enhanced Wi-Fi and OcuSync detection streams into a centralized web application for a comprehensive drone detection solution.

ACKNOWLEDGMENT

A special thanks goes to David Protzman for releasing the DJI OcuSync drone ID open-source project. Without that resource, the demodulation steps and logic would have been unknown. I would also like to thank Dr. Jason Staggs for his insightful comments, feedback and suggestions on making this research better.

REFERENCES

[1] 3rd Generation Partnership Project, Technical Specification, Group Radio Access Network, Evolved Universal Terrestrial Radio Access (E-UTRA), Physical Channels and Modulation (Release 10), 3GPP TS 36.211 V10.7.0., 2013.

[2] A. Chadd, The FreeBSD Project, Half and Quarter Rate Support, Boulder, CO (wiki.freebsd.org/dev/ath_hal%284%29/HalfQuarterRate), 2012.

[3] Cybersecurity and Infrastructure Security Agency, Unmanned Aircraft Systems (UAS): Critical Infrastructure, Arlington, Virginia (www.cisa.gov/uas-critical- infrastructure), 2021.

[4] Da-Jiang Innovations, areoscope_type, (mydjiflight.dji.com/links/links/areoscope_type), 2022.

[5] Da-Jiang Innovations, DJI AeroScope, Shenzhen, China (www.dji.com/aeroscope), 2022.

[6] Da-Jiang Innovations, DJI introduces voluntary flight identification options for drone pilots, DJI News, Shenzhen, China (www.dji.com/newsroom/news/dji-introduces-voluntary-flight-identification-options-for-drone-pilots), December 1, 2017.

[7] Da-Jiang Innovations, How to Check Your Product's Serial Number, (repair.dji.com/product/serial/index), 2022.

[8] Department 13, Anatomy of DJI's Drone Identification Implementation, White Paper, Canberra, Australia, 2017.

[9] Federal Aviation Administration, Aircraft Inquiry, Washington, DC (registry.faa.gov/aircraftinquiry), 2022.

[10] Federal Aviation Administration, Remote Identification of Unmanned Aircraft, Code of Federal Regulations, Title 14, Chapter 1, Part 89, Washington, DC, 2022.

[11] Federal Aviation Administration, Small Unmanned Aircraft Systems, Code of Federal Regulations, Title 14, Chapter 1, Part 107, Washington, DC, 2022.

[12] Federal Aviation Administration, UAS Remote Identification Overview, Washington, DC (www.faa.gov/uas/getting started/remote id), 2021.

[13] M. Gast, 802.11 Wireless Networks: The Definitive Guide, O'Reilly Media, Sebastopol, California, 2005.

[14] Heliguy, DJI transmission systems – Wi-Fi, OcuSync and Lightbridge, Newcastle upon Tyne, United Kingdom (www.heliguy.com/blogs/posts/dji-transmission- systems-wi-fi-ocusync-lightbridge), May 5, 2021.

[15] S. Hollister, DJI insisted drone-tracking AeroScope signals were encrypted — now it admits they aren't, The Verge, April 28, 2022.

[16] Institute of Electrical and Electronics Engineers, standards-oui, (standards-oui.ieee.org), 2022.

[17] MathWorks, Channel Model, (ww2.mathworks.cn/discovery/channel-model.html), Natick, MA, 2022.

[18] D. Protzman, DJI DroneID RF Analysis, GitHub (www.github.com/proto17/dji_droneid), 2022.

[19] T. Tsou, TurboFEC, GitHub, (https://github.com/ttsou/turbofec), 2018.

[20] J. Xianjun, LTE-Cell-Scanner, GitHub (www.github.com/JiaoXianjun/LTE-Cell-Scanner), 2022.

[21] F. Xu and H. Muneyoshi, A case study of DJI, the top drone maker in the world, Kindai Management Review, vol. 5, pp. 97–99, 2017.